\documentclass[11pt, a4paper, twocolumn]{aastex631}
\usepackage{acronym}
\usepackage{subfigure}
\usepackage{graphicx}
\usepackage{footmisc}
\usepackage{float}
\usepackage{tabularx}
\usepackage{appendix}
\usepackage{multirow}
\usepackage{ragged2e}
\acrodef{GW}[GW]{gravitational wave}
\acrodef{CNN}[CNN]{convolutional neural network}
\acrodef{EM}[EM]{electromagnetic}
\acrodef{FAP}[FAP]{false alarm probability}
\acrodef{FAR}[FAR]{false alarm rate}
\acrodef{FNP}[FNP]{false negative probability}
\acrodef{TAP}[TAP]{true alarm probability}
\acrodef{LVC}[LVC]{The LIGO Scientific and Virgo Collaborations}
\acrodef{OPA}[OPA]{Open Public Alert}
\acrodef{GCN}[GCN]{Gamma-ray Coordination Network}
\acrodef{BNS}[BNS]{binary neutron star}
\acrodef{CNN}[CNN]{convolutional neural network}
\acrodef{FITS}[FITS]{Flexible Image Transport System}
\acrodef{BBH}[BBH]{binary black hole}
\acrodef{NSBH}[NSBH]{neutron star black hole}
\acrodef{LVK}[LVK]{LIGO-Virgo-KAGRA Collaboration}
\acrodef{API}[API]{application programming interface}
\acrodef{LVC}[LVC]{LIGO Scientific and Virgo Collaboration}
\acrodef{O3}[O3]{the third observing run}
\acrodef{O2}[O2]{the second observing run}
\acrodef{O4}[O4]{the fourth observing run}
\acrodef{CBC}[CBC]{compact binary coalescence}
\acrodef{MDC}[MDC]{mock data challenge}
\acrodef{ROC}[ROC]{receiver operator characteristic}
\acrodef{SNR}[SNR]{signal-to-noise ratio}
\acrodef{O1}[O1]{the first}
\acrodef{O2}[O2]{second}
\acrodef{ER15}[ER15]{the engineering run}
\acrodef{GraceDB}[GraceDB]{the gravitational wave-candidate event DataBase}

\makeatletter
\newcommand\footnoteref[1]{\protected@xdef\@thefnmark{\ref{#1}}\@footnotemark}

\makeatother
\newcolumntype{Y}{>{\centering\arraybackslash}X}

\newcommand{\MC}[2][black]{\textcolor{#1}{#2}}
\begin{document}
\title{\large \bf \texttt{\bf GWSkyNet II} : a refined machine learning pipeline for real-time classification of public gravitational wave alerts}
\author[0009-0009-6826-4559]{Man Leong Chan}
\affiliation{Department of Physics and Astronomy, The University of British Columbia, Vancouver, BC V6T 1Z4, Canada}
\author[0000-0003-0316-1355]{Jess McIver}
\affiliation{Department of Physics and Astronomy, The University of British Columbia, Vancouver, BC V6T 1Z4, Canada}
\author[0000-0003-2242-0244]{Ashish Mahabal}
\affiliation{Division of Physics, Mathematics and Astronomy, California Institute of Technology, Pasadena, CA 91125, USA}
\affiliation{Center for Data Driven Discovery, California Institute of Technology, Pasadena, CA 91125, USA}
\author[0000-0002-8230-3309]{Cody Messick}
\affiliation{University of Wisconsin-Milwaukee, Milwaukee, WI 53201, USA}
\author[0000-0001-6803-2138]{Daryl Haggard}
\affiliation{Department of Physics, McGill University, 3600 rue University, Montreal, Quebec H3A2T8, Canada}
\affiliation{Trottier Space Institute at McGill, 3550 rue University, Montreal, Quebec H3A2A7, Canada}
\author[0000-0002-8549-9124]{Nayyer Raza}
\affiliation{Department of Physics, McGill University, 3600 rue University, Montreal, Quebec H3A2T8, Canada}
\affiliation{Trottier Space Institute at McGill, 3550 rue University, Montreal, Quebec H3A2A7, Canada}
\author[0000-0002-9186-7034]{Yannick Lecoeuche}
\affiliation{Department of Physics and Astronomy, The University of British Columbia, Vancouver, BC V6T 1Z4, Canada}
\author[0000-0003-1614-3922]{Patrick J. Sutton}
\affiliation{Gravity Exploration Institute, Cardiff University, Cardiff CF24 3AA, UK}
\author[0000-0001-9178-5744]{Becca Ewing}
\affiliation{Department of Physics, The Pennsylvania State University, University Park, PA 16802, USA}
\affiliation{Institute for Gravitation and the Cosmos, The Pennsylvania State University, University Park, PA 16802, USA}
\author[0000-0002-5447-3810]{Francesco Di Renzo}
\affiliation{Université Lyon, Université Claude Bernard Lyon 1, CNRS, IP2I Lyon/IN2P3, UMR 5822, F-69622 Villeurbanne, France}
\author[0000-0003-4059-4512]{Miriam Cabero}
\affiliation{Department of Physics and Astronomy, The University of British Columbia, Vancouver, BC V6T 1Z4, Canada}
\author{Raymond Ng}
\affiliation{Department of Computer Science, University of British Columbia, Vancouver, British Columbia, V6T1Z4, Canada}
\author[0000-0002-8262-2924]{Michael W. Coughlin}
\affiliation{School of Physics and Astronomy, University of Minnesota, Minneapolis, Minnesota 55455, USA}
\author[0000-0001-9901-6253]{Shaon Ghosh}
\affiliation{Montclair State University, 1 Normal Ave. Montclair, NJ 07043}
\author[0000-0002-7489-4751]{Patrick Godwin}
\affiliation{LIGO Laboratory, California Institute of Technology, Pasadena, CA 91125, USA}


\begin{abstract}
Electromagnetic follow-up observations of gravitational wave events offer critical insights and provide significant scientific gain from this new class of astrophysical transients. Accurate identification of gravitational wave candidates and rapid release of sky localization information are crucial for the success of these electromagnetic follow-up observations. However, searches for gravitational wave candidates in real time suffer a non-negligible false alarm rate. By leveraging the sky localization information and other metadata associated with gravitational wave candidates, \texttt{GWSkyNet}, a machine learning classifier developed by \cite{cabero2020gwskynet}, demonstrated promising accuracy for the identification of the origin of event candidates. We improve the performance of the classifier for LIGO-Virgo-KAGRA's fourth observing run by reviewing and updating the architecture and features used as inputs by the algorithm. We also retrain and fine-tune the classifier with data from the third observing run. To improve the prospect of electromagnetic follow-up observations, we incorporate \texttt{GWSkyNet} into LIGO-Virgo-KAGRA's low-latency infrastructure as an automatic pipeline for the evaluation of gravitational wave alerts in real time.
We test the readiness of the algorithm on a LIGO-Virgo-KAGRA mock data challenge campaign. 
The results show that by thresholding on the \texttt{GWSkyNet} score, noise masquerading as astrophysical sources can be rejected efficiently and the majority of true astrophysical signals  correctly identified.   
\vspace{1cm}
\end{abstract}
\section{\bf Introduction}
The \ac{LVK} has so far reported nearly a hundred \ac{GW} events from \acp{CBC}
~\citep{2019PhRvX...9c1040A, 2021PhRvX..11b1053A, 2021arXiv211103606T}.
Sources of \acp{GW} such as \acp{CBC} involving at least one neutron star may also produce accompanying \ac{EM} emissions. 
This was demonstrated by the first directly observed \ac{BNS} inspiral by the \ac{LVC}, referred to as \ac{GW}170817~\citep{PhysRevLett.119.161101}.
Simultaneous observations of the \ac{GW} event and the associated gamma-ray burst triggered an extensive multi-messenger campaign, which resulted in the detection of the associated counterparts across a wide range of X-ray through radio wavelength~\citep{doi:10.1126/science.aap9811, Smartt:2017fuw, goldstein2017ordinary, savchenko2017integral, Abbott_2017, LIGOScientific:2017zic}.

Coincident \ac{EM} and \ac{GW} observations of \acp{CBC} involving a neutron star clearly provide valuable information leading to a deeper understanding of the sources.
Such observations can help constrain the equation of state of neutron star matter~\citep{Radice:2017lry, Bauswein_2017, Margalit_2017, 10.1093/mnrasl/slz133, 10.1093/mnras/sty2174, 10.1093/mnras/stz3457, PhysRevLett.120.172703, PhysRevLett.120.261103, Lai:2018ugk, Dietrich:2020efo, huth2022constraining}, and provide insights into the central engine of short gamma-ray bursts~\citep{LIGOScientific:2017zic}. A better understanding of the origin of $r$-process elements in the Universe can also be achieved~\citep{drout2017light, doi:10.1126/science.aap9811, chornock2017electromagnetic, cowperthwaite2017electromagnetic, Pian:2017gtc, Rosswog:2016dhy, Smartt:2017fuw, Watson:2019xjv, 10.1093/mnrasl/slz007}. Additionally, such observations can break the modeling degeneracies of the source properties and independently measure the Hubble constant~\citep{LIGOScientific:2017adf, PhysRevResearch.2.022006, coughlin2020measuring, Hotokezaka:2018dfi, Dietrich:2020efo}.

However, the success of multi-messenger observations of \ac{GW} sources depends crucially on a number of factors including
the sky localization and the rapid release of the \ac{GW} alerts.
The $90\%$ credible region the size of $\mathcal{O}(10)$ to $\mathcal{O}(100)~\mathrm{deg}^2$ can be expected with a large network of \ac{GW} detectors operating together~\citep{KAGRA:2013rdx, petrov2022data, Kiendrebeogo:2023hzf, Pankow:2019oxl}.
Rapid release of \ac{GW} alerts, on the other hand, is accomplished with fast data analysis pipelines~\citep{Ewing:2023qqe, Chu:2020pjv, PhysRevD.98.024050, dal2021real, adams2016low, Aubin:2020goo}.  

During \ac{O3}, \ac{GW} public alerts\footnote{\MC{https://emfollow.docs.ligo.org/}\label{userguidefootnote}} were released on the Gamma-ray Coordinates Network (GCN)\footnote{\MC{https://gcn.nasa.gov/}} within minutes of identification by low-latency \ac{GW} search pipelines, which have built-in tests for the consistency of the phase, amplitude, and timing of candidates in each detector.
In \ac{O3}, a total of $80$ alerts were published by the \ac{LVK}. Of these alerts, however, $24$ were subsequently retracted and another $11$ were not reported in the offline analyses~\citep{2021PhRvX..11b1053A, 2021arXiv211103606T}. Some of these alerts had already triggered electromagnetic follow-up observations such as S191213g and S200213t~\citep{10.1093/mnras/staa1846}. This underscores the need for a model capable of reliably determining the origin of \ac{GW} alerts in real time.


In 2020, \citet{cabero2020gwskynet} developed \texttt{GWSkyNet}, a machine learning classifier capable of separating \ac{GW} events from detector noise transients. The classifier was trained with simulated \ac{GW} signals from \acp{CBC} and non-Gaussian noise transients, known as ``glitches", from \texttt{Gravity Spy} collected by the \ac{LVC} during \ac{O1} and \ac{O2} observing runs~\citep{2017CQGra..34f4003Z, bahaadini2018machine, 2019PhRvD..99h2002C, Zevin:2023rmt}.
Assuming \ac{GW} events with $p_{astro}> 0.5$ as reported by~\cite{2021arXiv211103606T, LIGOScientific:2021usb} are of astrophysical origin,  the classifier 
correctly identified 59 of the 77 \ac{CBC} public alerts in \ac{O3} (assuming a \texttt{GWSkyNet} score threshold of 0.75, see section~\ref{sec:implement}), showing promising scientific potential~\citep{cabero2020gwskynet}. Inspired by the success of \texttt{GWSkyNet}, \citet{2021arXiv211104015A} built an extension to the classifier known as \texttt{GWSkyNet-Multi}, for the identification of the presence or absence of a neutron star in the source system in addition to noise versus signal classifications. \texttt{GWSkyNet-Multi} was reported to show accuracy comparable to \texttt{GWSkyNet}~\citep{2021arXiv211104015A}. Further study has been undertaken to understand and reveal the inner workings of \texttt{GWSkyNet-Multi}~\citep{nayyer2023}. The results will better inform  development of the algorithm and generation of more effective and representative training data.

Given the performance of \texttt{GWSkyNet}, the scientific potential of the classifier can be maximized by implementing it as a real-time pipeline for the classification of \ac{GW} alerts to facilitate potential \ac{EM} follow-up observations for \ac{O4}. 
To improve the performance of \texttt{GWSkyNet} for \ac{O4}, 
we retrain and fine-tune the classifier using the data collected by the \ac{LVK} during \ac{O3}. 
The inputs from a \ac{GW} candidate to the classifier and the classifier's architecture are reviewed and updated to be compatible with the alert data expected for public alerts in \ac{O4}.\footnoteref{userguidefootnote} 
To demonstrate the readiness of \texttt{GWSkyNet}, we participate in a \ac{MDC}~\citep{MDC_paper} campaign organized by the \ac{LVK}.

The paper is organized as follows: In section~\ref{sec:GWSkyNet}, we provide an overview of the latest changes made to \texttt{GWSkyNet}.
This includes a discussion of the development of the classifier since its first publication in~\citet{cabero2020gwskynet}.  
In section~\ref{sec:performance}, we show the readiness and performance of \texttt{GWSkyNet} based on the \ac{LVK}-led \ac{MDC} campaign.
We discuss in section~\ref{sec:implement} the deployment and implementation of \texttt{GWSkyNet} as an automatic pipeline in \ac{LVK}'s low-latency infrastructure. 
A summary is given in section~\ref{sec:sum}. 
\section{\bf \texttt{GWS\lowercase{ky}N\lowercase{et}}: A machine learning classifier}\label{sec:GWSkyNet}
\texttt{GWSkyNet}, developed by \citet{cabero2020gwskynet}, is a machine learning classifier with a non-sequential \ac{CNN} architecture. \acp{CNN} are deep learning algorithms commonly used for image classification. In \ac{GW} astronomy, deep learning algorithms have been applied to the identification of signals and noise~\citep{Gabbard:2017lja, Chan:2019fuz}, parameter estimation~\citep{Gabbard:2019rde, Green:2020hst}, sky localization~\citep{chatterjee2022rapid} and detector noise transient generation~\citep{Powell:2022pcg} (for a review, see \cite{Cuoco_2021}). Compared to more traditional methods for \ac{GW} data analysis, an advantage of deep learning algorithms is that they are relatively computationally cheap to run while the performance is comparable~\citep{heaton2018ian}. 

\texttt{GWSkyNet} is designed to distinguish between noise transients and astrophysical signals by leveraging sky localization information and meta data associated with a \ac{GW} candidate. The features of the data that \texttt{GWSkyNet} takes as inputs are mixed, including images and meta data. This allows \texttt{GWSkyNet} to evaluate a \ac{GW} candidate by weighing relevant information with different modalities.
\subsection{Inputs and Outputs}\label{sec:trainingfeatures}
For \ac{CBC} public alerts identified in real time, the \ac{LVK} generates \ac{FITS} files containing the sky localization information within seconds to minutes of the identification of the alerts\footnoteref{userguidefootnote} using \texttt{BAYESTAR}~\citep{Singer:2015ema, Singer_2016}. 
Information from the \ac{FITS} files that is important for the identification of the origin of a \ac{GW} candidate is extracted and used as input to the \texttt{GWSkyNet} model.
A complete list of the inputs  for the current \texttt{GWSkyNet} models (Baseline and Fine-tuned models, see sections~\ref{sec:retraining} and~\ref{sec:fineting}) is shown in the right column of Table~\ref{table:table1}.
Also listed in Table~\ref{table:table1} are the inputs used by the original \texttt{GWSkyNet} model presented in~\cite{cabero2020gwskynet} for comparison.
\begin{table*}[]
\centering
\begin{tabular}{lcccc}
\toprule
                                  &\multicolumn{4}{c}{Features}                                                                     \\
\hline

                                  && \citet{cabero2020gwskynet} &            & Baseline and Fine-tuned models  \\ 
\hline
\multirow{4}{*}{Image inputs}     && 2D sky map ($90 \times 180$)                 && 2D sky map ($180 \times 360$)   \\
                                  && \texttt{DISTMU} ($90 \times 180$)            && \texttt{Vol0} ($262 \times 262$) \\
                                  && \texttt{DISTSTD}($90 \times 180$)            && \texttt{Vol1} ($262 \times 262$)\\
                                  && \texttt{DISTNORM}($90 \times 180$)           && \texttt{Vol2} ($262 \times 262$)\\
\hline                              
\multirow{8}{*}{Numerical inputs} && A binary array with 3 elements for detectors\textsuperscript{\textdagger}     && A binary array with 3 elements for detectors\textsuperscript{\textdagger} \\
                                   && Normalizing factor for the \texttt{DISTMU} image\textsuperscript{\textdaggerdbl}     && Normalizing factor for the \texttt{Vol0} image\textsuperscript{\textdaggerdbl} \\
                                   && Normalizing factor for the \texttt{DISTSTD} image\textsuperscript{\textdaggerdbl}    && Normalizing factor for the \texttt{Vol1} image\textsuperscript{\textdaggerdbl}\\
                                   && Normalizing factor for the \texttt{DISTNORM} image\textsuperscript{\textdaggerdbl}    && Normalizing factor for the \texttt{Vol2} image\textsuperscript{\textdaggerdbl}\\
                                   && Normalizing factor for the 2D sky map image\textsuperscript{\textdaggerdbl}         && Normalizing factor for the 2D sky map image\textsuperscript{\textdaggerdbl}  \\
                                   && Normalized posterior mean distance   && Normalized posterior mean distance  \\
                                   &&                           && Normalized \texttt{log BCI}               \\
                                   &&                           && Normalized \texttt{log BSN}                  \\
\hline             
\end{tabular}
\raggedright\footnotesize{{\textsuperscript{\textdagger}For \cite{cabero2020gwskynet} and the Baseline model, the elements are used to indicate whether the LIGO Hanford, LIGO Livingston, and Virgo detectors were operating at the time of a \ac{GW} candidate. For the Fine-tuned model, however, the elements are used to indicated the detectors with SNR $\geq 4.5$. Future updates of GWSkyNet will extend the model to incorporate KAGRA~\citep{10.1093/ptep/ptaa125, PhysRevD.88.043007} and LIGO India~\citep{KAGRA:2013rdx, iyerligo}.}}\\
\vspace{0.1cm}
\footnotesize{{\textsuperscript{\textdaggerdbl}All normalizing factors are also normalized by the maximum of the corresponding value from the entire training data set.}}
\caption{The inputs for the \texttt{GWSkyNet} model. The first column indicates whether the corresponding input feature is images or numerical data. 
The second and third columns show the inputs to the \texttt{GWSkyNet} model published in~\cite{cabero2020gwskynet} and those for the updated \texttt{GWSkyNet} model presented in this paper respectively. The numbers in the brackets indicate the image dimension in pixels. \label{table:table1}}
\end{table*}

\begin{figure*}
    \centering
    \setlength\tabcolsep{-10pt}
    \subfigure[Vol0]{%
        \includegraphics[width=0.656\columnwidth]{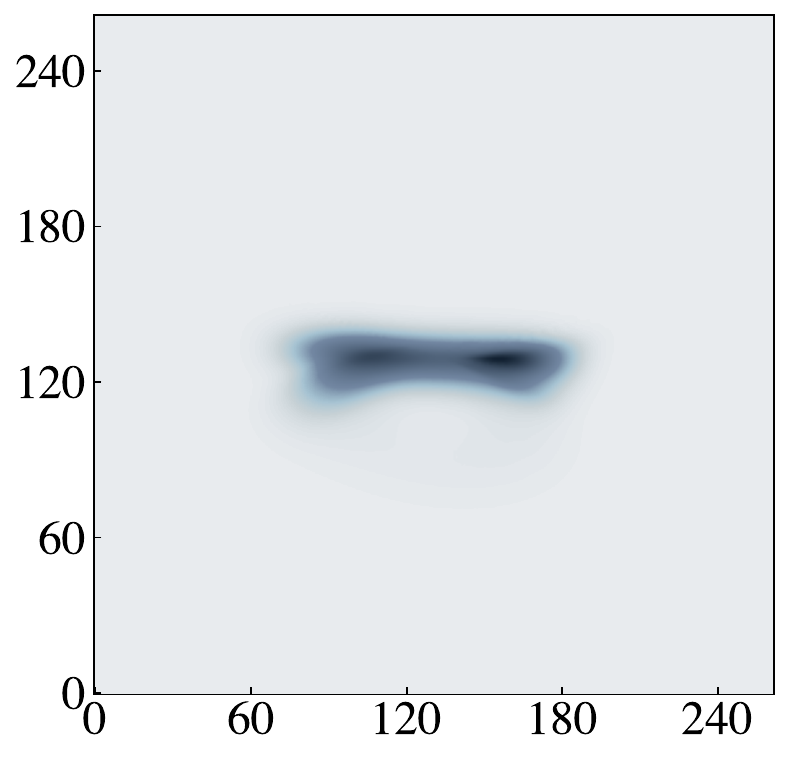}
        \label{fig:vol0}
    }%
    \subfigure[Vol1]{%
        \includegraphics[width=0.6\columnwidth]{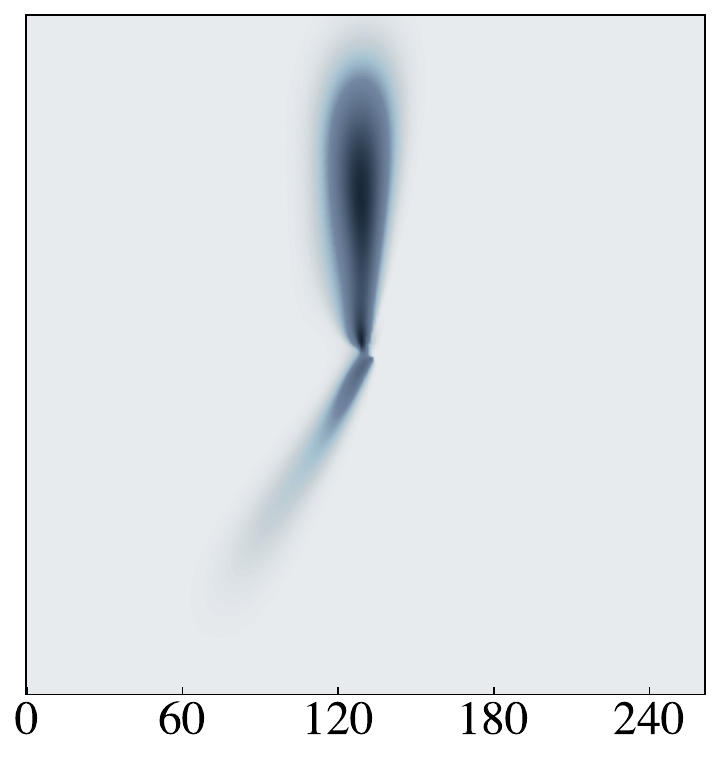}
        \label{fig:vol1}
   }%
   \subfigure[Vol2]{%
        \includegraphics[width=0.6\columnwidth]{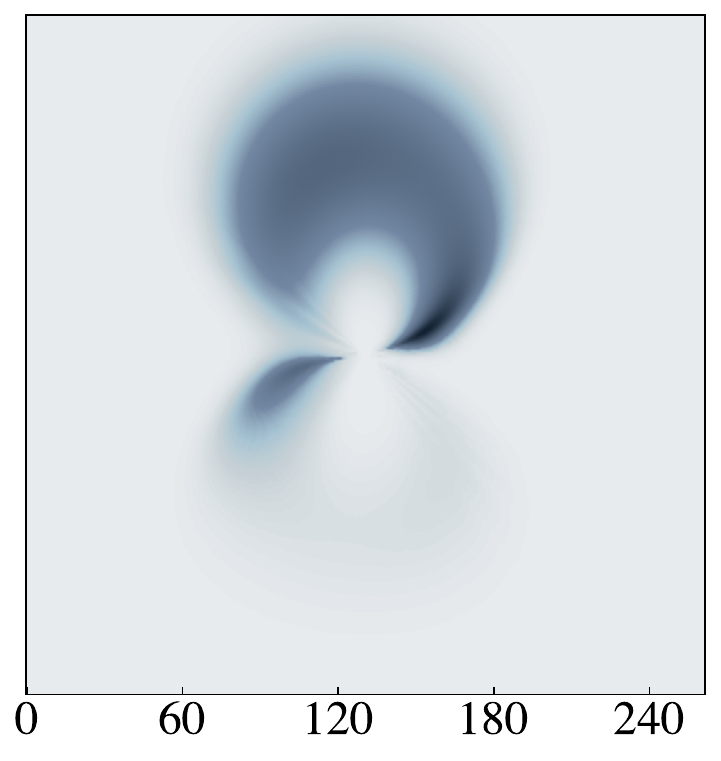}
        \label{fig:vol2}
   }\quad
   \hspace*{-1.75cm}\quad
   \subfigure[Sky localization]{%
        \includegraphics[width=1.3\columnwidth]{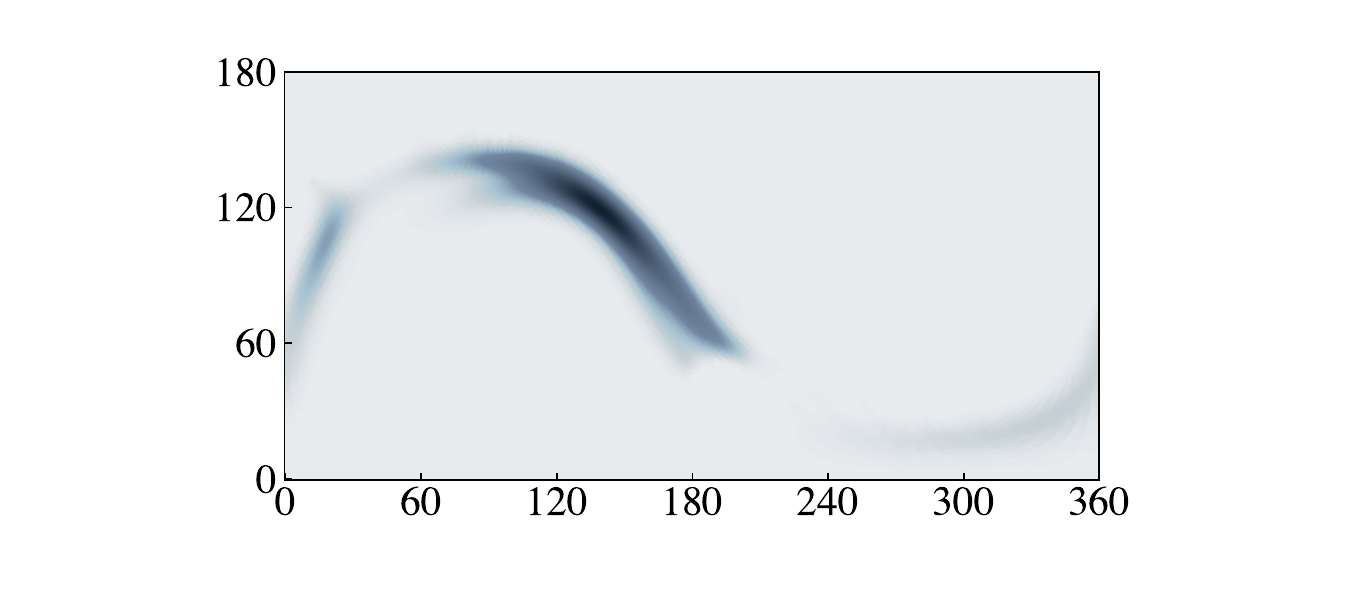}
        \label{fig:skymap}
   }
   \hspace*{-1.4cm}\quad
    \subfigure[Numerical data]{%
        \includegraphics[width=0.92\columnwidth]{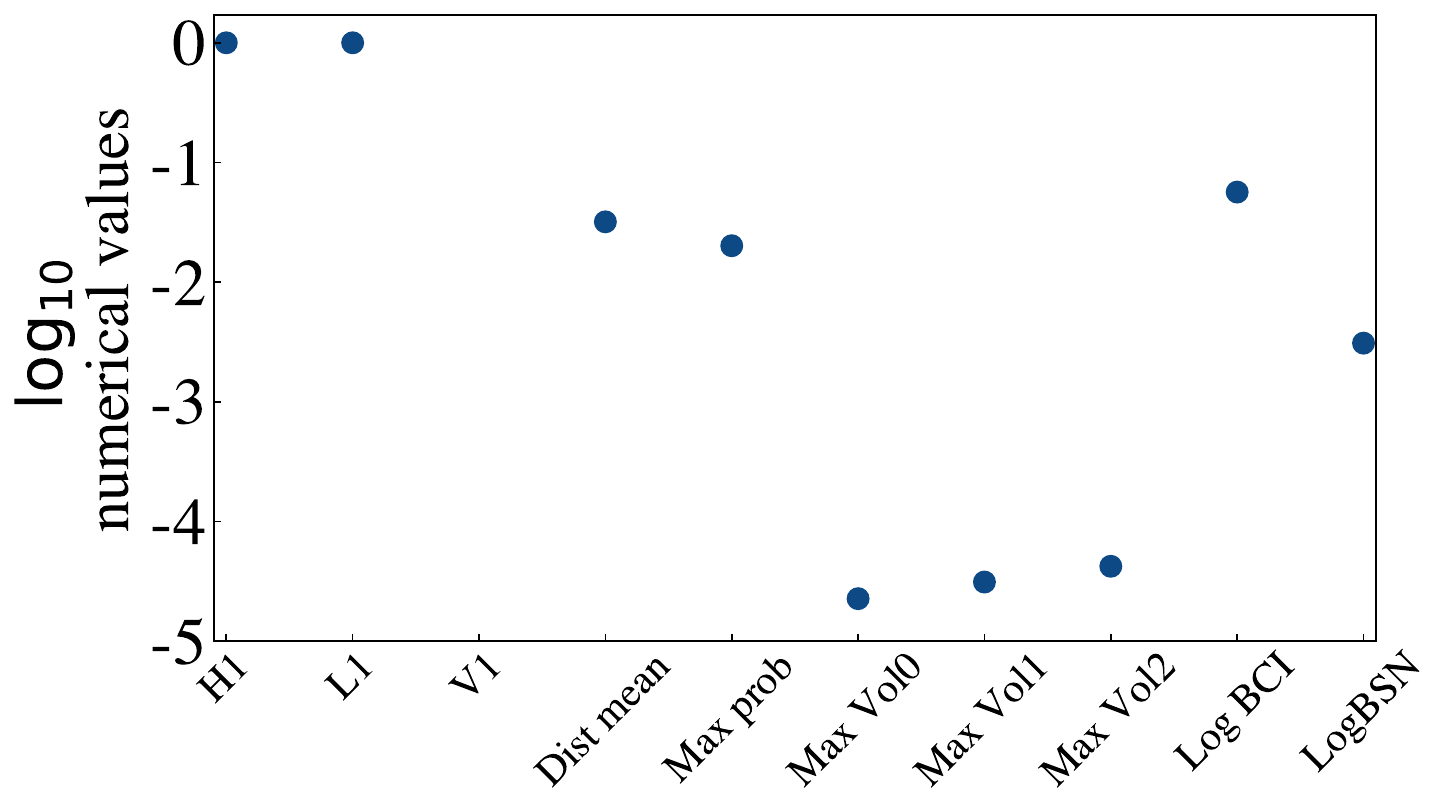}
        \label{fig:numericaldata}
   }
\caption{Input images and numerical data extracted from the \ac{FITS} file of S191120at for \texttt{GWSkyNet}. The top row shows the 3 volume reconstruction images of the event. 
The left plot in the bottom row shows a 2D projection of the sky localization. For the images, the x and y axes indicate the pixels. The bottom right plot shows the numerical data extracted from the associated \ac{FITS} file that are also used as inputs.\label{fig:exampleinputs}}
\end{figure*}

\texttt{GWSkyNet} is designed to be a multiple input model consisting of two \acp{CNN} and multiple multi-layer perceptrons. Each branch of the ensemble is dedicated to the analysis of a different feature from the input data. The outputs of the branches are concatenated together to form a final prediction. The current architecture of \texttt{GWSkyNet} is similar to that shown in~\cite{cabero2020gwskynet}. Several changes are made to the inputs to allow the model to learn a more accurate generalization.
First, as stated in~\cite{cabero2020gwskynet}, sky localization of event candidates is extracted and transformed to a 2D projection with $180\times360$ pixels.
However, the 2D projection is no longer down-sampled to $90\times180$ using maximum pooling.
This allows the classifier to learn and perform the down-sampling of the sky localization projection. 
In addition, instead of utilizing the ansatz location \texttt{DISTMU}, the scale parameters \texttt{DISTSTD} and the normalization coefficient \texttt{DISTNORM}
as inputs for space volume localization information, volume reconstruction images are employed. Volume reconstruction images are 2D projections of 3D probability volume map on three planes perpendicular to the principle components of the probability distribution, known as volume rendering images~\citep{Singer_2016}.
The change is made in response to the fact that the values of the normalization coefficient \texttt{DISTNORM} from \ac{FITS} files generated by more recent versions of \texttt{BAYESTAR} can be up to $10^{50}$, creating increased difficulty for data normalization. Using volume rendering images to describe probability volume map is also consistent with the way public alerts are shown on \ac{GraceDB}.\footnote{https://gracedb.ligo.org/} For the binary array for the presence and absence of detectors, we adopt a more refined physical definition to better capture event candidate property~(see section~\ref{sec:fineting}). Two other quantities not previously consumed by \texttt{GWSkyNet}, the log of the Bayes factors for coherence versus incoherence (\texttt{log BCI}) and signal versus noise (\texttt{log BSN}) from the \ac{FITS} files are also used as inputs.  As an example, the inputs extracted from the \ac{FITS} file of a public alert, S191120aj, released in \ac{O3} are shown in Figure~\ref{fig:exampleinputs}.

Other input information such as the posterior mean 
distance, the maximum value of \ac{GW} posterior probability in the sky localization and the normalizing factors for 
each of the input images are used by both versions of \texttt{GWSkyNet}. 

By virtue of being a binary classifier, the output from \texttt{GWSkyNet} is a class score ranging from 0 (consistent with noise) to 1 (consistent with astrophysical signal) indicating the model's confidence that the input localization information is internally consistent and aligns with that of a \ac{GW} signal. We call this score the \texttt{GWSkyNet} score for the remainder of the paper. \MC{\texttt{GWSkyNet} is applicable as long as the sky localization information is
produced by \texttt{BAYESTAR}. Currently, this is true for \texttt{PyCBC}~\citep{PhysRevD.98.024050, dal2021real}, \texttt{GstLAL}~\citep{PhysRevD.95.042001, PhysRevD.108.043004, PhysRevD.109.042008},
\texttt{SPIIR}~\citep{PhysRevD.105.024023} and \texttt{MBTA}~\citep{Aubin_2021}.}
\subsection{Retraining}\label{sec:retraining}
Optimal performances of machine learning models are possible only if they are trained on data representative of those expected in practice. Applying a machine learning model to data from a distribution different from that of the data it has been trained on may result in sub-optimal performances.
As different noise sources may contribute differently to the noise budget across observing runs, data from separate observing runs may show distinctive features and characteristics. For example, glitches may have diverse morphology across observing runs because instrument configurations have changed. Similar glitches may also occur at different rates. 

To ensure the classifier learns a generalization that is representative of the \ac{GW} candidates expected in \ac{O4}, we retrain and fine-tune the \texttt{GWSkyNet} model. This section will focus on retraining and the next section on fine-tuning. To retrain, we first compile training samples 
using \ac{O3} data from the Gravitational Wave Open Science Centre.\footnote{https://gwosc.org/} Since \texttt{GWSkyNet} is a signal versus noise classifier, we generate data of these two classes.
For both classes, we employ a procedure similar to that presented in~\citet{cabero2020gwskynet}. In particular, we simulate a population of \acp{GW} from mergers of \acp{BBH}, \acp{NSBH} and \acp{BNS} using gravitational waveform models available in the LALSuite package~\citep{lalsuite}. 
To allow the simulated data to be as realistic and representative of those expected in low latency as possible, we \MC{add the simulated \acp{GW} to Gaussian noise colored by} the power spectral densities associated with the \ac{GW} alerts released during \ac{O3} from \ac{GraceDB}. 
The simulated signals are distributed equally among these different representative power spectral densities.

A population of noise events from known glitches in \ac{O3} is also identified for training using a \ac{GW} trigger list from the fourth Open Gravitational-wave Catalog~\citep{Nitz:2021zwj} and a collection of glitches from the \texttt{Gravity Spy} classifier. 
\texttt{Gravity Spy} is a scientific initiative that combines human expertise, machine learning and citizen scientist with the aim of classifying detector noise transients based on their morphologies~\citep{2017CQGra..34f4003Z, bahaadini2018machine, 2019PhRvD..99h2002C, Zevin:2023rmt}. 

\begin{figure}
\includegraphics[width=0.50\textwidth]{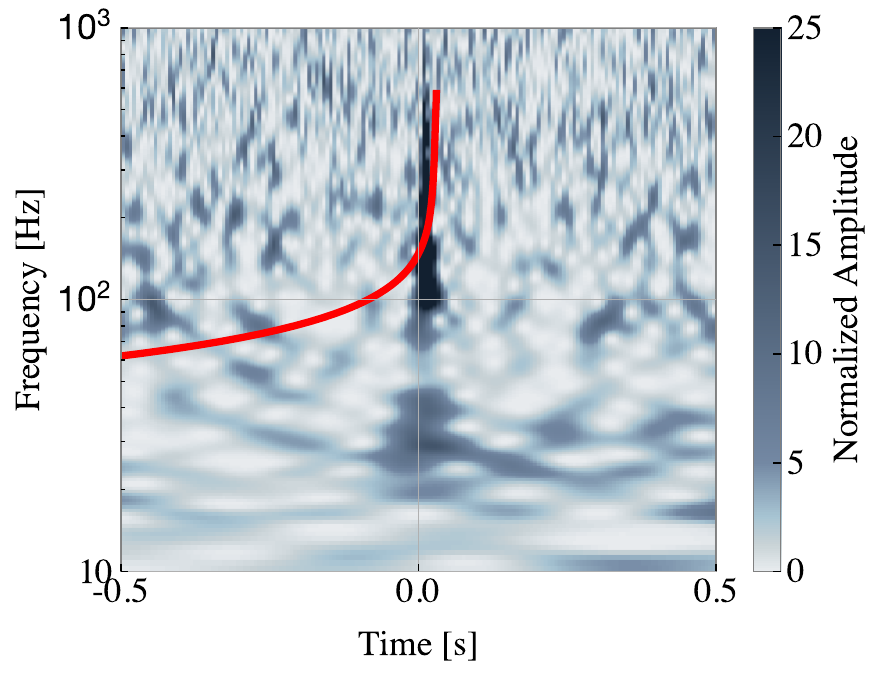}
\caption{Time-frequency representation of a blip glitch at LIGO Livingston from \ac{O3}. 
The red curve represents the \ac{GW} template that triggered the 4-OGC search. 
This template corresponds to a \ac{NSBH} merger with component masses of $m_{1}=23M_{\odot}$ and $m_{2}=2M_{\odot}$. 
The network \ac{SNR} of the candidate by the search is $7.6$.
\label{fig:noiseglitchexample}}
\end{figure}

A \ac{SNR} requirement is then implemented to select the candidates 
that are observed in at least two detectors with an \ac{SNR} of no less than $4.5$ and a network \ac{SNR}, given by the quadrature sum of the individual \ac{SNR}, of greater than or equal to $7.0$. In total, $\sim17400$ event candidates were generated for the training of the \texttt{GWSkyNet} model, with approximately $8700$ for each class. \MC{In particular, the $\sim8700$ remaining noise events include all types of glitches found in \texttt{Gravity Spy}.}
An example noise candidate is shown in Figure~\ref{fig:noiseglitchexample}. A gravitational waveform of a coincident 4-OGC trigger is also shown. %
A model trained on this data set will be able to learn features and patterns in \ac{GW} candidates broadly consistent with those expected in low-latency searches, allowing to establish a baseline performance. For the rest of the paper, this training data set will be referred to as the Baseline data set and the model as the Baseline model.
\subsection{Fine-tuning}\label{sec:fineting}
To evaluate the performance of the \ac{LVK} low-latency infrastructure and its readiness for \ac{O4}, the \ac{LVK} collaboration has organized 
a \ac{MDC} campaign as part of the preparation for O4. The \ac{MDC} consisted of $40$ days of detector data collected by LIGO and Virgo from 
5 Jan. 2020 15:59:42 to 14 Feb. 2020 15:59:42 UTC. A total of  $5 \times 10^4$ simulated \ac{GW} signals from 
mergers of \acp{BNS}, \acp{NSBH} and \acp{BBH} with a minimum network \ac{SNR} of 4 were added to the data at a rate of one per $\sim 40$ seconds. 
The data were then replayed and analyzed by search pipelines in low-latency configurations as if the data were just collected. Once the replay of the 40 days of data is completed, a new cycle will ensue. Details regarding the distributions of the simulated \ac{GW} signals can be found in~\cite{MDC_paper}. The \ac{MDC} campaign allows search pipelines to refine their configurations for low-latency searches such as the construction of template banks, the ranking statistics and the estimation of candidate significance~\citep{Ewing:2023qqe}, leading to more accurate and sensitive search pipelines.

However, such evolving configurations could result in identifying a population of \ac{GW} candidates that differs from that represented by the Baseline data set. In addition, the \ac{FAR} estimate of \ac{GW} candidates by search pipelines in low-latency is a key statistic in determining whether the candidates will be released as public alerts,\footnoteref{userguidefootnote} while such information is generally not available for the Baseline data set.

To overcome the weaknesses, we use the \ac{GW} candidates identified in a \ac{MDC} cycle of the \ac{MDC} campaign by search pipelines to fine-tune the Baseline model to \ac{GW} candidates generated by low-latency searches. Fine-tuning is a common technique in transfer learning. During fine-tuning, the weights of a pre-trained model are fine-tuned on a target data set (see~\cite{pan2009survey}). As only the event candidates with a \ac{FAR} $\leq 2$ per day will be released as public alerts in \ac{O4},\footnoteref{userguidefootnote} we first select all the noise event candidates with \ac{FAR} $\leq 2$ per day. The \ac{SNR} requirement defined in section~\ref{sec:retraining} is applied, leaving approximately $300$ candidates. An equal number of simulated \ac{GW} events are then selected randomly. \MC{We then augment the data by shifting the observing time randomly at which the \ac{GW} candidates were observed, changing the right ascension and declination of the sky localizations. This results in a data set of $\sim12000$ samples.} This approaches the maximum number achievable with augmentation by changing the observing times, as shifting the observing time by multiples of approximately 24 hours would lead to an identical sky localization. We will refer to this data set as the Fine-tuned data set. Since the number of distinct \ac{GW} candidates in this data set is significantly lower than that in the Baseline data set, allowing all the layers of the Baseline model to be trainable may result in the model forgetting the generalization learned from the Baseline data set and overfitting the Fine-tuned data set. We therefore freeze the majority of the layers of the Baseline models and only allow the three dense layers in the two convolutional branches and the final layer to be trainable (see Figure 3 in~\cite{cabero2020gwskynet}). Also, we find that by adopting a more refined physical definition for the input binary array for detectors to indicate detectors with a \ac{SNR} of $4.5$ or greater  (see Table~\ref{table:table1}), a better performance can be achieved. This is because data from all three detectors may be used to produce \ac{GW} candidate localization even if the candidate has negligible \ac{SNR} in one detector. The resulting model is referred to as the Fine-tuned model. Both the Baseline model and the Fine-tuned model are then applied to a new and more recent cycle of the \ac{MDC} campaign to evaluate the performance, which will be referred to as the test \ac{MDC} cycle. The results are shown in the next section.

\section{\bf Performance}\label{sec:performance}
\begin{figure}
    \centering
    \subfigure[]{%
        \includegraphics[width=\columnwidth]{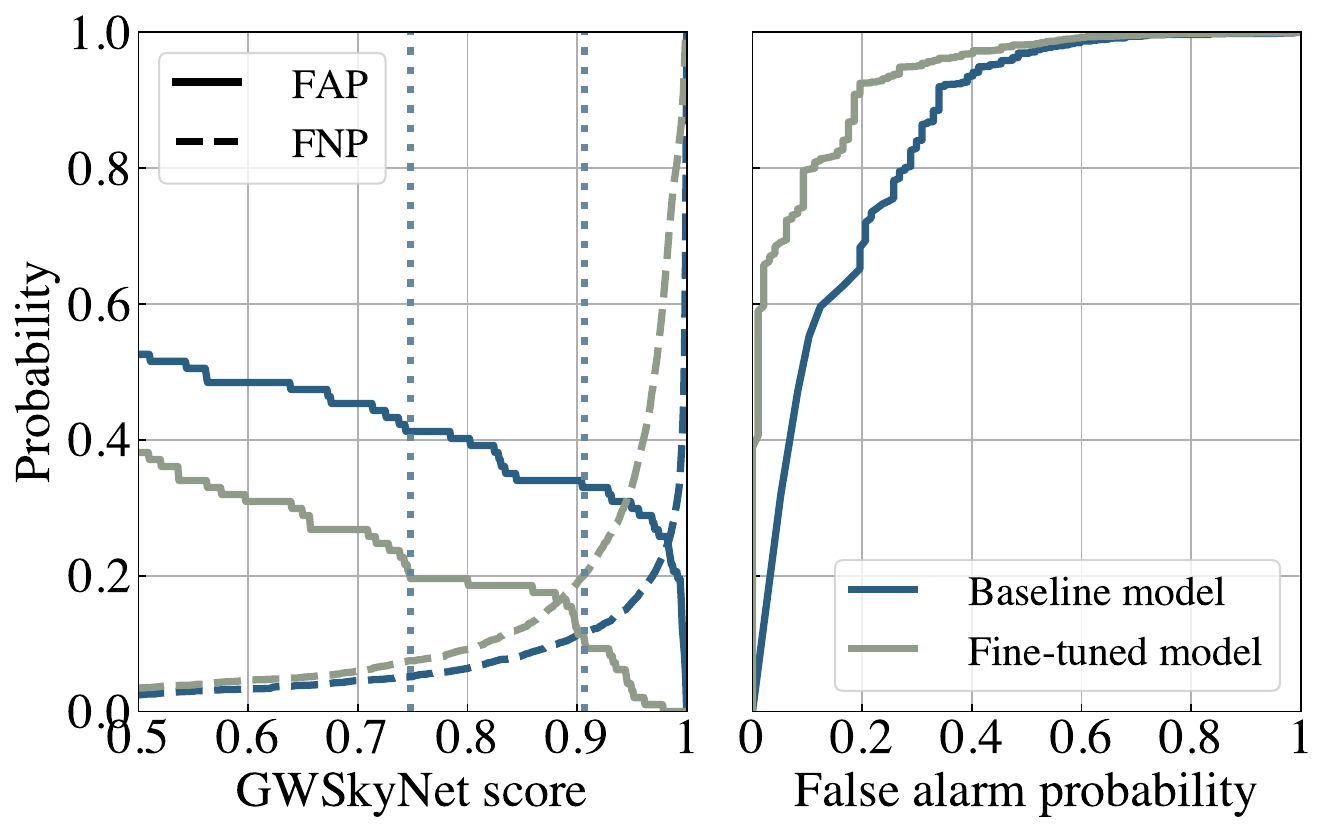}
        \label{fig:scorevsprob}
    }\qquad
    \subfigure[]{%
        \includegraphics[width=\columnwidth]{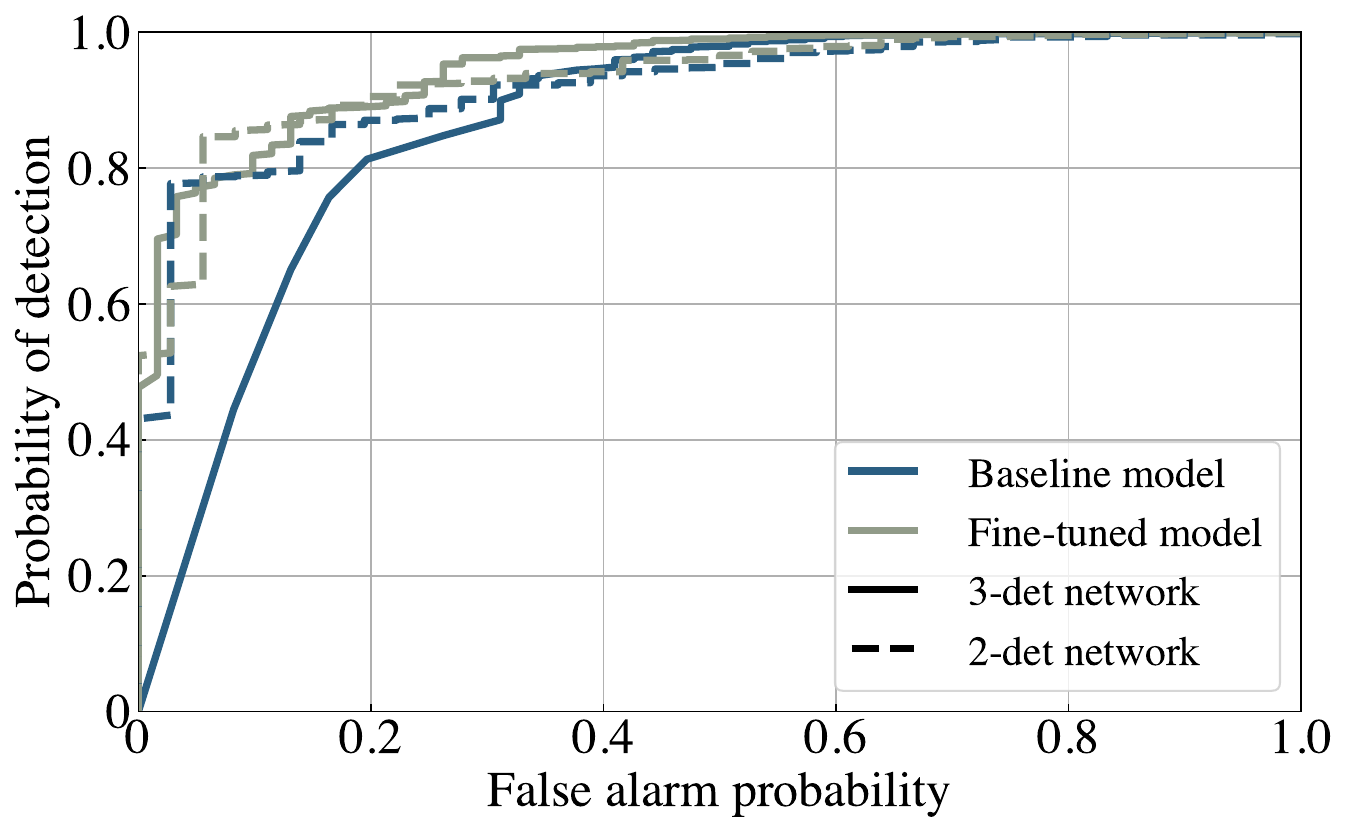}
        \label{fig:better_performance}
   }
\caption{The performance of both the Baseline and Fine-tuned model on the \ac{LVK}-led \ac{MDC} (the test \ac{MDC} cycle). The upper left panel shows the
\ac{FAP} and \ac{FNP} as a function of the \texttt{GWSkyNet} score, while the upper right panel shows the \MC{probability of detection} as a function of
\ac{FAP}. The dotted lines in the upper left panel indicate the \acp{FAP} and \acp{FNP} at two different \texttt{GWSkyNet} scores corresponding 
to two different use cases with the Fine-tuned model described in section~\ref{sec:implement}. 
The upper panels are for all events that met the \ac{SNR} requirement specified in section~\ref{sec:retraining}.
The bottom panel shows the performances of the Baseline and Fine-tuned models for events identified by two-detector networks and three-detector networks respectively.
\label{fig:results}}
\end{figure}

During the test \ac{MDC} cycle, $4016$ event candidates were identified by at least one \ac{LVK} \ac{CBC} search pipeline with a \ac{FAR} $\leq2$ per day that met the \ac{SNR} requirement specified in section~\ref{sec:retraining}. Of these $4016$ candidates, $3913$ were simulated \ac{GW} events and $96$ were noise transients. 
The remaining $7$ were real \ac{GW} events reported in~\cite{2021arXiv211103606T}.
We quantify the performance of both the Baseline model and the Fine-tuned model on these $4016$ events with \ac{FAP} and \ac{FNP}. The \ac{FAP} is an empirical measure of the fraction of noise event candidates at or above a candidate’s \texttt{GWSkyNet} score. Similarly, the \ac{FNP} is an empirical measure of the fraction of simulated \ac{GW} candidates at or below a candidate's \texttt{GWSkyNet} score. In the upper left panel of Figure~\ref{fig:results} we show the \ac{FAP} and \ac{FNP} as a function of \texttt{GWSkyNet} score.

\begin{figure}
    \centering
    \subfigure[]{%
        \includegraphics[width=\columnwidth]{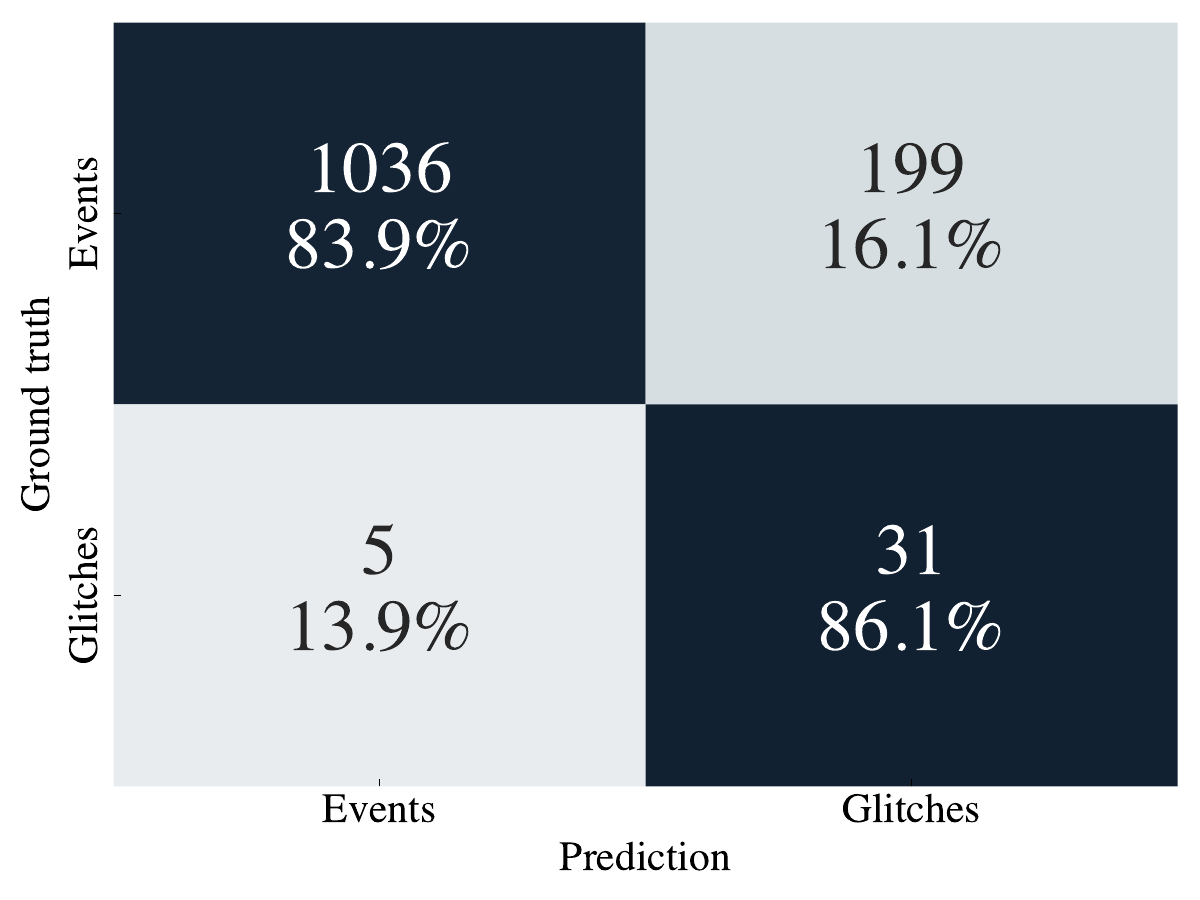}
        \label{fig:bl_cm2d}
    }\qquad
    \subfigure[]{%
        \includegraphics[width=\columnwidth]{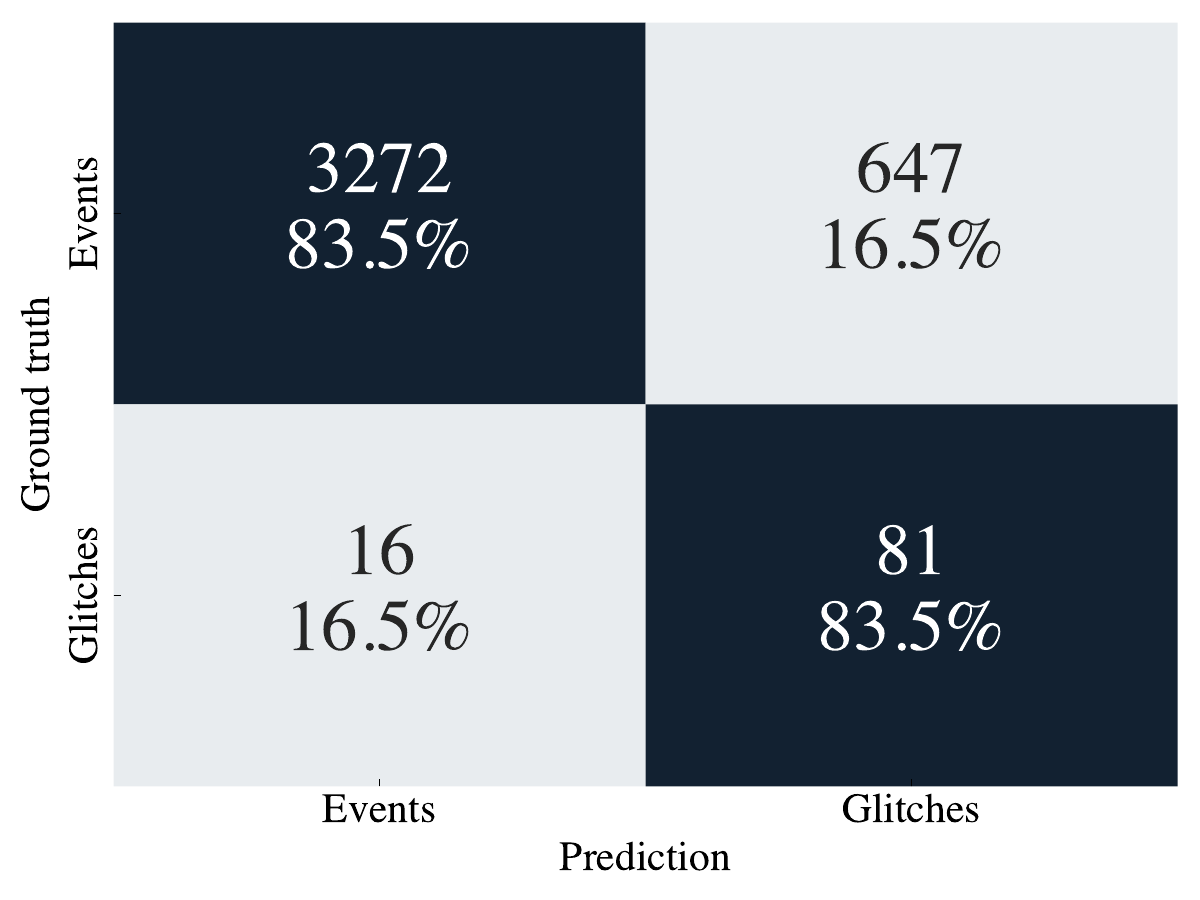}
        \label{fig:ft_cmall}
   }
\caption{Confusion matrices depicting the performances of the Baseline and Fine-tuned models on events identified during the test \ac{MDC} cycle.
The upper panel shows the Baseline model's performance for two-detector events, and the lower shows the Fine-tuned model's performance for both two-detector and three-detector events. The confusion matrices are generated using a \texttt{GWSkyNet} score threshold that balances the \ac{FAP} and \ac{FNP}. For the Baseline model, this threshold is $0.785$, while it is $0.885$ for the Fine-tuned model.
\label{fig:cm}}
\end{figure}

To present the performance of \texttt{GWSkyNet} from another perspective, we also construct the \ac{ROC} curve in the upper right panel. 
The \ac{ROC} curve is one of the most commonly used and convenient ways
to determine the classification performance of a signal detection algorithm. It shows the
performance of a classifying model by defining the \MC{probability of detection}, which is the fraction of simulated GW candidates correctly identified, as a function of the \ac{FAP}. 
The \ac{ROC} curve is obtained by computing the \ac{FAP} and \MC{probability of detection} at different values of \texttt{GWSkyNet} score.
For a given \ac{FAP}, a model with a higher \MC{probability of detection} is considered more capable than a model with a lower \MC{probability of detection}. 
At a \ac{FAP} of $20\%$, the Fine-tuned model achieves a \MC{probability of detection} of $92.5\%$ while the Baseline model obtains a \MC{probability of detection} of $68.3\%$ clearly suggesting 
the necessity of fining-tuning \texttt{GWSkyNet} to the configurations of search pipelines. However, we note that the improvement in the performance is
largely attributed to an increase in \MC{probability of detection} at \ac{FAP} $\lesssim 0.35$ for three-detector events. As shown in Figure~\ref{fig:better_performance}, 
for event candidates identified by networks of two detectors, the Baseline and Fine-tuned models have comparable performance. For example, at a \ac{FAP} of $20\%$, the \acp{TAP} for the Baseline and the Fine-tuned models are $91\%$ and $88\%$ for two detector events, while the \acp{TAP} are $90\%$ and $81\%$ for three-detector events.
By adjusting the \texttt{GWSkyNet} score threshold, one can achieve a desired \MC{probability of detection}, while keeping an acceptable \ac{FAP} based on one's risk tolerance. The confusion matrices in Figure~\ref{fig:cm} demonstrate the performance achieved by thresholding on a \texttt{GWSkyNet} score so that \ac{FAP} and \ac{FNP} are balanced. We expect \ac{O4} will first operate with a two-detector configuration and transition to a three-detector configuration with Virgo joining in early 2024.\footnoteref{userguidefootnote} This will coincide with
the expected implementation of the Baseline and Fine-tuned models (see section~\ref{sec:implement}). Therefore, for the Baseline model, we show the performance for two-detector events at a \texttt{GWSkyNet} score threshold of $0.785$, while for the Fine-tuned model, the performance is for all events at a \texttt{GWSkyNet} score threshold of $0.885$. 
However, the small number of glitches available in the test \ac{MDC} cycle suggests the performance of the models may deviate in practice. The results can be taken only as reference. 

\begin{figure}
\includegraphics[width=0.46\textwidth]{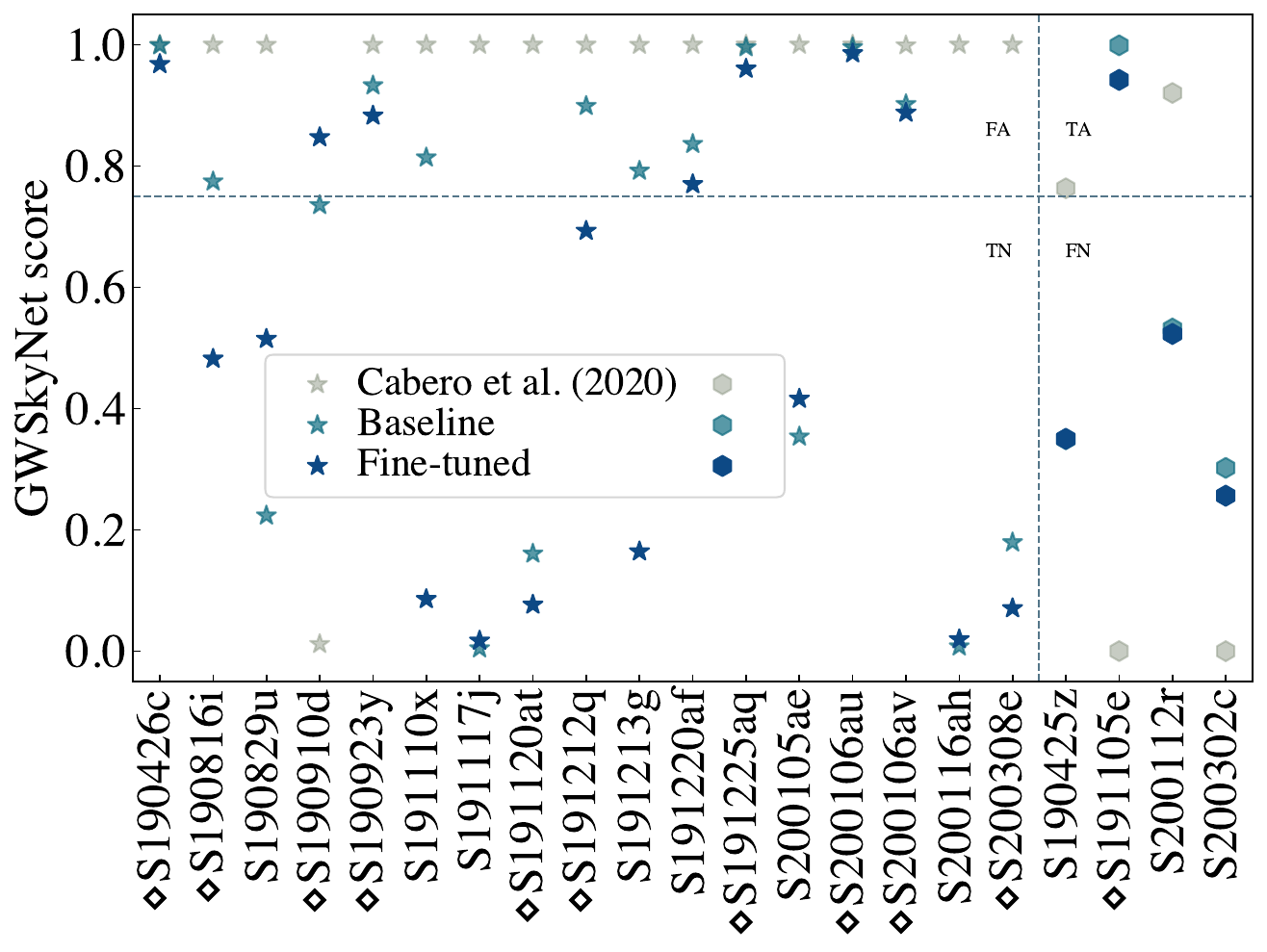}
\caption{A scatter plot showing the misidentifications of O3 alerts by the GWSkyNet models (Baseline model, Fine-tuned model and \cite{cabero2020gwskynet}. The horizontal axis shows the IDs of the alerts that are misclassified by at least one model. The symbol $\diamond$ next to the IDs indicates whether the \ac{SNR} of the alerts would have met the \ac{SNR} requirement defined in section~\ref{sec:retraining} and thus whether would have been analyzed by \texttt{GWSkyNet}. The vertical axis is the \texttt{GWSkyNet} score. Stars indicate glitches and hexagons indicate \ac{GW} signals according to \cite{2021arXiv211103606T, LIGOScientific:2021usb}. The GWSkyNet score threshold (0.75) used to determine the misidentifications is indicated by the horizontal dashed line. The vertical dashed line separate glitches and \ac{GW} events.
\label{fig:o3mis}}
\end{figure}

In addition, to allow for a direct comparison between the model in~\cite{cabero2020gwskynet} and the models in this paper, we apply both the Baseline and Fine-tuned model to the $77$ \ac{CBC} alerts published in \ac{O3}. \MC{To ensure this is a representative test, we use the sky localization information produced by
\texttt{BAYESTAR} that was immediately available at the time of the alerts.} Of the $77$ alerts, $64$ and $67$ are correctly identified by the Baseline and Fine-tuned models respectively (assuming a \texttt{GWSkyNet} score threshold of 0.75 (see section~\ref{sec:implement}) and according to~\cite{2021arXiv211103606T, LIGOScientific:2021usb}), which is $5$ and $8$ more correct identifications or $\sim6\%$ and $\sim10\%$ more accurate than the original \texttt{GWSkyNet} model, showing improvement over an already well performing model. Figure~\ref{fig:o3mis} shows the misclassifications by the models. The misclassifications are also presented in a tabular format in Table~\ref{table:o3table} in Appendix. As a reference, we also provide the \texttt{GWSkyNet} scores by the Baseline model (see section~\ref{sec:implement}) for \ac{O4} public alerts in Table~\ref{table:table2}. \MC{For $95\%$ of the O4 public alerts analyzed
by \texttt{GWSkyNet}, the time between the release of the public alerts and the completion of the analysis by \texttt{GWSkyNet} is $\leq 66$ seconds.} 
\begin{table*}[!t]
\centering
\begin{tabularx}{\textwidth}{XXp{0.2cm}}
\toprule
Alert ID  & GWSkyNet score & Retracted \\
\hline
S230518h  & 0.995          & False     \\
S230520ae & 0.995          & False     \\
S230601bf & 0.931          & False     \\
S230605o  & 0.986          & False     \\
S230606d  & 0.937          & False     \\
S230608as & 0.981          & False     \\
S230609u  & 0.978          & False     \\
\textbf{S230622ba} & \textbf{0.611}          & \textbf{True}      \\
S230624av & 0.970          & False     \\
S230627c  & 0.987          & False     \\
S230628ax & 0.993          & False     \\
S230630am & 0.991          & False     \\
S230630bq & 0.933          & False     \\
S230702an & 0.967          & False     \\
S230704f  & 0.863          & False     \\
S230706ah & 0.978          & False     \\
S230707ai & 0.962          & False     \\
S230708t  & 0.950          & False     \\
S230708z  & 0.981          & False     \\
S230708cf & 0.934          & False     \\
S230709bi & 0.970          & False     \\
\textbf{S230715bw} & \textbf{0.000}          & \textbf{True}      \\
S230723ac & 0.948          & False     \\
S230729z  & 0.987          & False     \\
S230731an & 0.992          & False     \\
S230805x  & 0.972          & False     \\
S230806ak & 0.975          & False     \\
S230807f  & 0.921          & False     \\
S230811n  & 0.994          & False     \\
S230814r  & 0.992          & False     \\
S230819ax & 0.988          & False     \\
S230820bq & 0.695          & False     \\
S230824r  & 0.992          & False     \\
S230825k  & 0.935          & False     \\
S230831e  & 0.989          & False	   \\
S230904n  & 0.989          & False     \\
S230914ak & 0.997          & False     \\
S230919bj & 0.999          & False     \\
S230920al & 0.988          & False     \\
S230922g  & 0.895          & False     \\
S230922q  & 0.997          & False     \\
S230924an & 0.997          & False     \\
S230927l  & 0.995          & False     \\
S230927be & 0.997          & False     \\
S230928cb & 0.999          & False     \\
S230930al & 0.965          & False     \\
S231001aq & 0.942          & False     \\
S231005j  & 0.999          & False     \\
S231005ah & 0.967          & False    \\
\hline 
\end{tabularx}
\caption{The \texttt{GWSkyNet} scores by the Baseline model for the \ac{O4} \ac{GW} candidates that have been published as of this writing, \MC{using the sky localization information that was immediately available at the time of the alerts.} 
The list is not comprehensive as only the alerts where the \ac{SNR} meet the \ac{SNR} requirements defined in section~\ref{sec:retraining} were analyzed with \texttt{GWSkyNet}. Assuming a \texttt{GWSkyNet} score threshold of 0.75 (see section~\ref{sec:implement}), \texttt{GWSkyNet} has correctly classified the retracted events (S230622ba and S230715bw). \label{table:table2}}
\end{table*}

\subsection{Interpretation when signals and glitches overlap}
\begin{figure}
    \centering
    \subfigure[]{%
        \includegraphics[width=1\columnwidth]{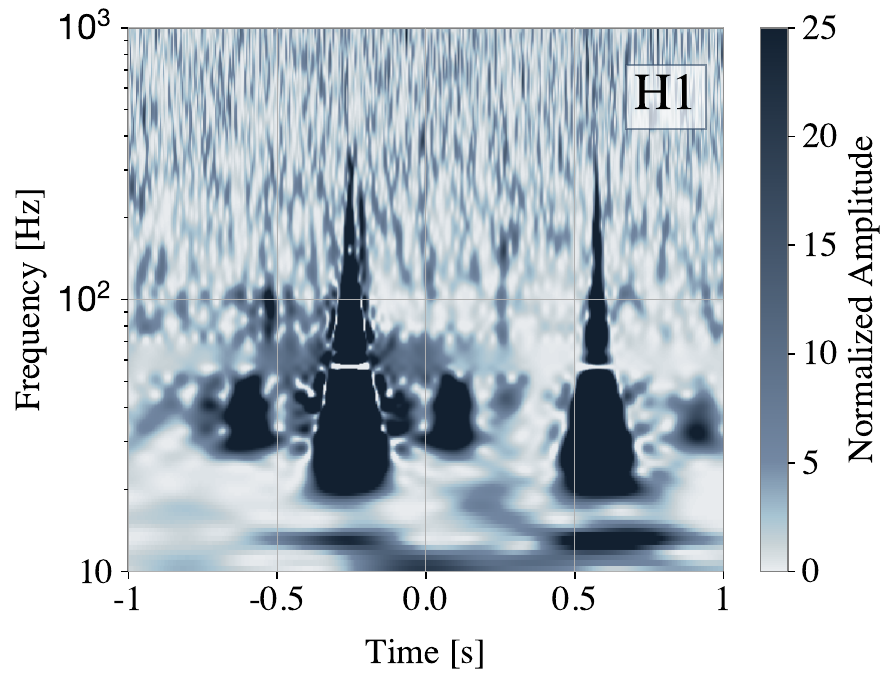}
        \label{fig:h1_G874295}
    }\qquad
    \subfigure[]{%
        \includegraphics[width=1\columnwidth]{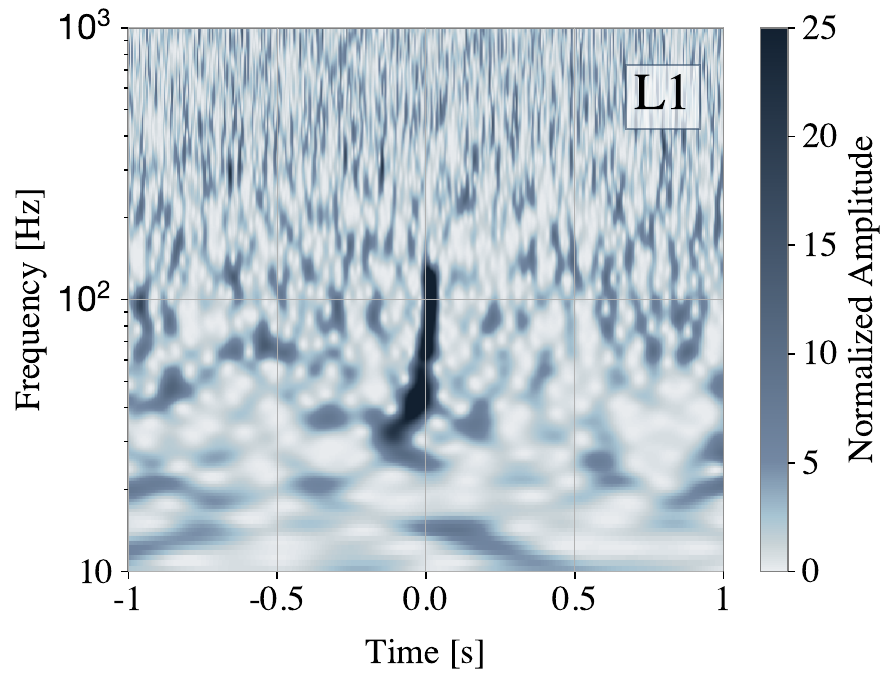}
        \label{fig:l1_G874295}
   }\qquad
   \subfigure[]{%
        \includegraphics[width=1\columnwidth]{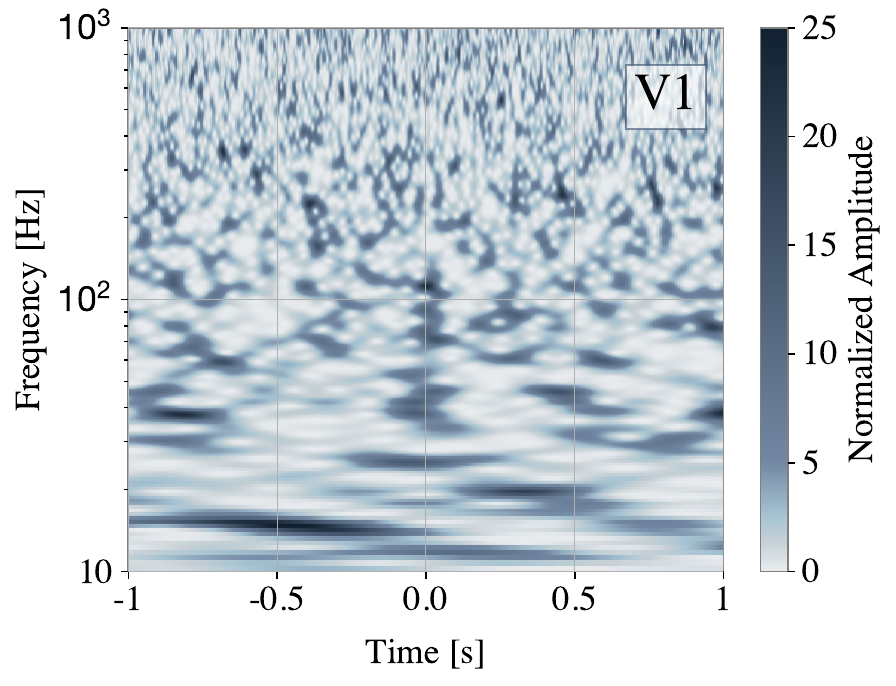}
        \label{fig:v1_G874295}
   }
\caption{Time-frequency representations of a simulated \ac{GW} candidate in the LIGO and Virgo detectors. The merger time of the simulated \ac{GW} is at 0\,s. The signal is visible in the spectrogram for LIGO Livingston (b). However, a loud glitch is coincident with the \ac{GW} signal and dominates the spectrogram for LIGO Hanford (a).
\label{fig:omega_scan_G874295}}
\end{figure}

\begin{figure}
\centering
    \subfigure[]{%
        \includegraphics[width=1\columnwidth]{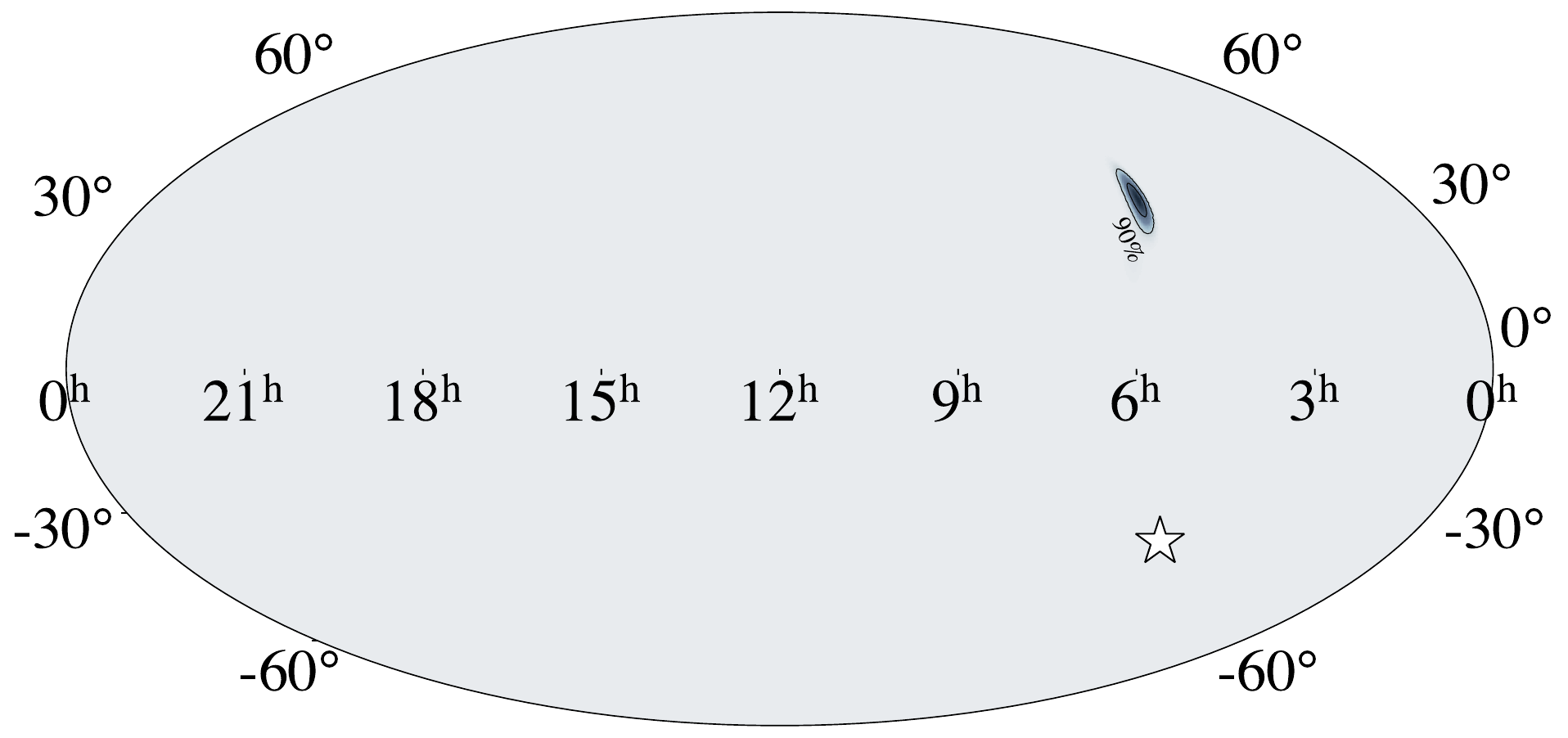}
        \label{fig:skymap_G874295}
    }\qquad
    \subfigure[]{%
        \includegraphics[width=1\columnwidth]{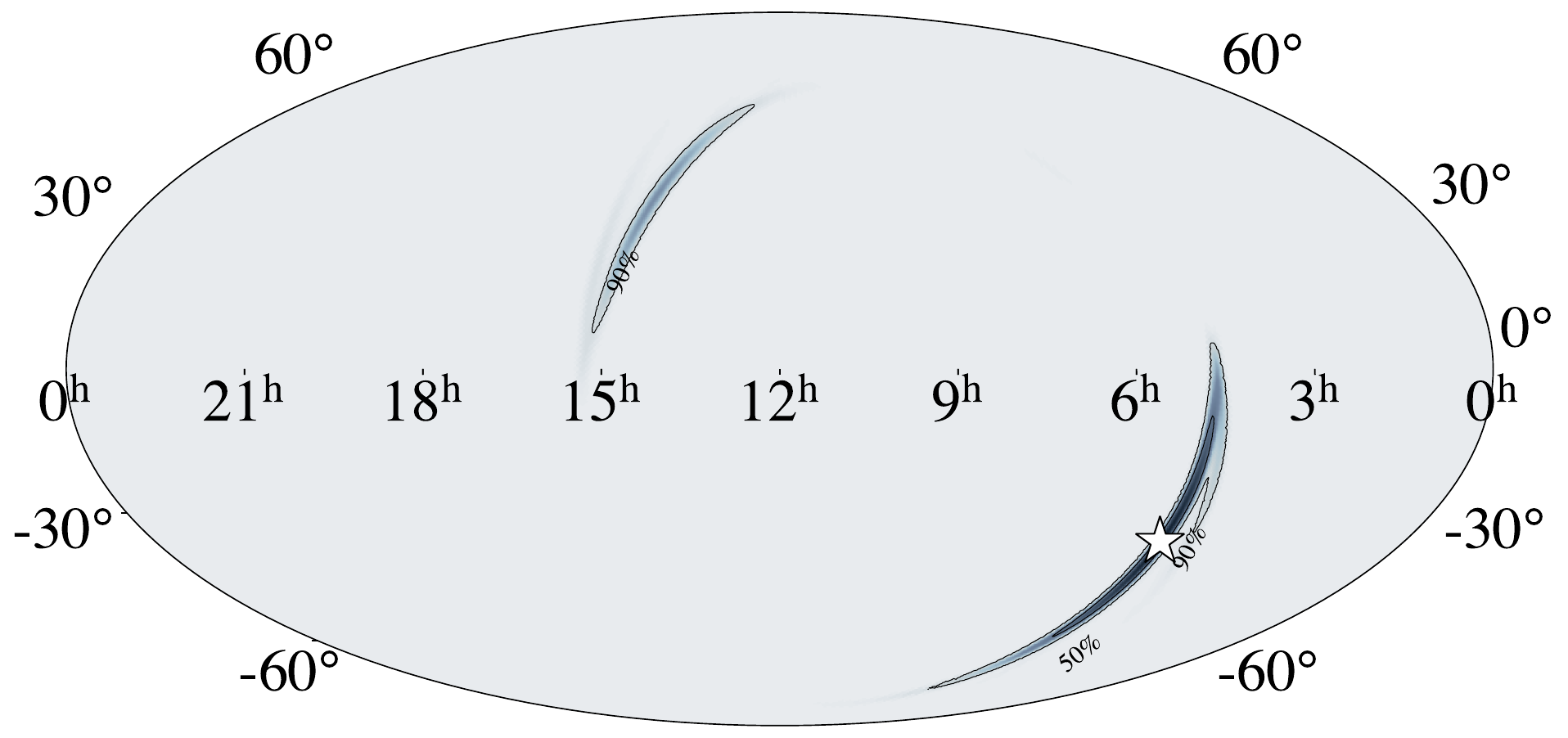}
        \label{fig:skymap_G874292}
    }
\caption{The sky localization of the simulated \ac{GW} event shown in Figure~\ref{fig:omega_scan_G874295} by \texttt{BAYESTAR}. 
The upper panel shows the localization by LIGO Hanford, LIGO Livingston and Virgo, while the lower shows that by LIGO Livingston and Virgo.
The true location of the event is marked by a star in both sky maps. In this example, the true location is not consistent with the $90\%$ credible region in the upper panel.
\label{fig:S230204gd_skymap}}
\end{figure}

Due to the high glitch rate of \ac{O3}~\citep{2021PhRvX..11b1053A, 2021arXiv211103606T}, a significant number of simulated \ac{GW} signals in the \ac{MDC} happened to be added at a time close to the occurrence of a glitch. The presence of a nearby or overlapping glitch may impact the estimates of the source parameters and the localization of the event candidate~\citep{Macas:2022afm, Huber:2022kbf, Pankow:2018qpo, Powell:2018csz}. An example is shown in Figure~\ref{fig:omega_scan_G874295}. A simulated \ac{GW} signal visible in LIGO Livingston is added at a time coincident with a glitch in LIGO Hanford. Two sky localization maps corresponding to the simulated \ac{GW} signals are shown in Figure~\ref{fig:S230204gd_skymap}. The upper panel
shows the sky localization by LIGO Hanford, LIGO Livingston and Virgo, while the lower panel shows that by LIGO Livingston and Virgo only.
The true location of the event is not consistent with the $90\%$ credible region in the upper panel because of the presence of the glitch. 
The \texttt{GWSkyNet} score by the Fine-tuned model for the upper and lower sky localization is $2.0\times10^{-13}$ and $0.96$ respectively. 
A low \texttt{GWSkyNet} score for an initial sky localization of a \ac{GW} candidate similar to the upper panel may be mistakenly assumed to mean the absence of a \ac{GW} signal in the data. Even if a more accurate sky map for the same event later becomes available, follow-up observations may still be discouraged if the sky map is produced with lower \ac{SNR} such that it is not evaluated by \texttt{GWSkyNet}.
Conversely, a high \texttt{GWSkyNet} score (e.g., above $0.90$) for a candidate where the source is located outside the $90\%$ credible region may lead to an assumption that the localization is accurate such that the source is located in the $90\%$ credible region. 
As stated above, the \texttt{GWSkyNet} score should be considered to mean the \texttt{GWSkyNet} model’s confidence that the localization information is both internally consistent and aligns with that of a \ac{GW} signal. Careful interpretation is necessary.
\section{\bf Implementation}\label{sec:implement}
\texttt{GWSkyNet} is an algorithm that analyzes the end products from search pipelines (i.e., the sky localization information encoded in \ac{FITS} files and the associated metadata). This allows \texttt{GWSkyNet} to evaluate \ac{GW} events from a perspective that is different from standard search pipelines, which usually analyze \ac{GW} detector time series data. \texttt{GWSkyNet} thus supplies a complementary metric for the interpretation of the origin of event candidates that can be used to reduce false positives attributed to detector noise and consistently compared across different search pipelines. The use of \texttt{GWSkyNet} also enables the identification of the sky maps for a \ac{GW} candidate more consistent with a \ac{GW} signal.

\ac{EM} counterparts of \ac{GW} events may fade quickly after merger and early epoch observations of \ac{EM} counterparts are critical in constraining the properties of \ac{GW} sources~\citep{Abbott_2017, drout2017light, doi:10.1126/science.aap9811, Smartt:2017fuw, goldstein2017ordinary, savchenko2017integral, Abbott_2017, LIGOScientific:2017zic}. For instance, optical and ultraviolet observation of the kilonova associated with a \ac{BNS} merger in the first few hours after merger is necessary for breaking emission model degeneracy~\citep{Arcavi:2018mzm}. For this purpose, \texttt{GWSkyNet} is most useful if implemented as an automatic low-latency pipeline that evaluates the origin of \ac{GW} alerts. To this end, we integrated \texttt{GWSkyNet} with the Baseline model into the \ac{LVK}'s low-latency infrastructure (\texttt{GWCelery}).\footnote{https://rtd.igwn.org/projects/gwcelery/en/latest/index.html} The Fine-tuned model will be implemented in our next update. This will coincide with Virgo joining \ac{O4} when event candidates may trigger three detectors simultaneously. The integration allows \texttt{GWSkyNet} to listen and react to public \ac{GW} alerts in real time. \texttt{GWSkyNet} is currently operating for \ac{O4}.

For a \ac{GW} public alert identified with a network \ac{SNR} of $\geq 7.0$ and an \ac{SNR} of $\geq 4.5$ in at least two detectors, a \texttt{GWSkyNet} score, and the empirical measurements of \ac{FAP} and \ac{FNP} given by Figure~\ref{fig:scorevsprob} are provided on \ac{GraceDB}. 
These three quantities allow astronomers and other \ac{GW} analyses to screen noise based on their tolerance of false positives and the amount of resources that they can spend. As indicated in Figure~\ref{fig:scorevsprob}, using the Fine-tuned model as an example, for an astronomer with ample telescope time who seeks to compile a list of promising candidates as large as possible, a \texttt{GWSkyNet} score
threshold of $0.75$ can be used to reject $80\%$ noise and capture $93\%$ astrophysical signals. On the other hand, if the astronomer is hoping to follow up on the \ac{GW} candidates most likely to be of astrophysical in origin, a \texttt{GWSkyNet} score threshold of $0.91$
can be used to reject $90\%$ noise and catch $80\%$ of astrophysical signals. However, we emphasize the small number of glitches in the test \ac{MDC} cycle 
suggests that these statistics can only be used as references and performance in practice may deviate. In the future, as \texttt{GWSkyNet} continues to operate for \ac{O4} and accumulates sufficient statistics, the \ac{FAP} and \ac{FNP} will be more reliable and can be incorporated into the \texttt{GWSkyNet} score
to provide a calibrated probability that a \ac{GW} candidate is of astrophysical origin.


\section{\bf summary}\label{sec:sum}
\texttt{GWSkyNet} is a machine learning classifier developed to facilitate \ac{EM} follow-up observations of \ac{GW} candidates. The classifier leverages the sky localization information of \ac{LVK} event candidates encoded in the \ac{FITS} file and the metadata associated with \ac{GW} candidates. The use of sky localization information instead of detector time series data allows a unique evaluation of event candidates. It also provides a metric that can be compared directly and interpreted consistently across all search pipelines. 

We reviewed and improved the architecture of the \texttt{GWSkyNet} model and the inputs 
to allow a more effective normalization of the data and better generalization by the classifier. 
Our results for the \ac{LVK} \ac{MDC} campaign show promising accuracy. In particular, using a \texttt{GWSkyNet} score threshold of $0.75$, the Fine-tuned model allows the rejection of more than $80\%$ noise while capturing more than $93\%$ astrophysical signals. 
Additionally, the Fine-tuned model correctly identify $67$ public alerts during \ac{O3} assuming a \texttt{GWSkyNet} score threshold of $0.75$, $8$ more than the original \texttt{GWSkyNet} model.

We also show that for signals where a glitch is close in time or overlapping, resulting in a noise-influenced sky localization, careful interpretation of the \texttt{GWSkyNet} score is encouraged. We emphasize the \texttt{GWSkyNet} score should be interpreted as the \texttt{GWSkyNet} model's confidence that the input localization information is consistent internally and with a \ac{GW} signal. 

To facilitate \ac{EM} follow-up observations of \ac{GW} events, we integrated \texttt{GWSkyNet} into \ac{LVK}'s low-latency infrastructure that responds to \ac{GW} candidates in real time in \ac{O4}. For each event candidate, three quantities are provided and made available on \ac{GraceDB}: \texttt{GWSkyNet} score and the empirical measurement of \ac{FAP} and \ac{FNP}. The use of these quantities allows a trade-off between rejecting noise and capturing true astrophysical signals.

\section*{Acknowledgment}
We are thankful to the following individuals for their valuable contributions to the development and success of \texttt{GWSkyNet}, without which GWSkyNet would not have been possible. 
We wish to thank Maria Drout, Aaron Tohuvavohu, Derek Davis, Brandon Piotrzkowski, Deep Chatterjee, Leo Singer, Erik Katsavounidis, Patrick Brockill, Reed Essick, Beverly Berger, Alan Knee and Jonah Kanner for their valuable feedback and suggestions, which helped us refine \texttt{GWSkyNet} as a pipeline. 
J.M. acknowledges funding from the Natural Sciences and Engineering Research Council of Canada (NSERC). A.M. acknowledges support from the NSF (1640818, AST-1815034). A.M. and J.M. also acknowledge support from IUSSTF (JC-001/2017). D.H. and J.M. acknowledge support from the Canada Research Chairs (CRC) program, the NSERC Discovery Grant program, and the Canadian Institute for Advanced Research (CIFAR). The authors highlight support for this project from the Canadian Tri-Agency New Frontiers in Research Fund – Exploration program. M.~W.~Coughlin acknowledges support from the National Science Foundation with grant numbers PHY-2347628 and PHY-2117997. N.R. acknowledges funding support from the Trottier Space Institute at McGill and the Walter C. Sumner Memorial Fellowship. This material is based upon work supported by NSF's LIGO Laboratory which is a major facility fully funded by the National Science Foundation.
\newpage
\appendix
\begin{table*}[!t]
\centering
\begin{tabularx}{\textwidth}{p{2.0cm}cYYYYp{5.1cm}}
\toprule
Alert ID  & \cite{cabero2020gwskynet} & Baseline & Fine-tuned & True label & \multicolumn{1}{c}{\ac{SNR} criteria met} \\
\hline
S190425z\textsuperscript{\textdaggerdbl \textasteriskcentered}  & 0.76   & 0.35     & 0.35       & Real       & False                     \\
S190426c\textsuperscript{\textdagger \textdaggerdbl \textasteriskcentered}  & 1.00   & 1.00     & 0.97       & Glitch     & True                      \\
S190816i\textsuperscript{\textdagger \textdaggerdbl}  & 1.00   & 0.77     & 0.48       & Glitch     & True                      \\
S190829u\textsuperscript{\textdagger}  & 1.00   & 0.22     & 0.51       & Glitch     & False                     \\
S190910d\textsuperscript{\textasteriskcentered}  & 0.01   & 0.74     & 0.85       & Glitch     & True                      \\
S190923y\textsuperscript{\textdagger \textdaggerdbl \textasteriskcentered}  & 1.00   & 0.93     & 0.88       & Glitch     & True                      \\
S191105e\textsuperscript{\textdagger}  & 0.00   & 1.00      & 0.94       & Real       & True                      \\
S191110x\textsuperscript{\textdagger \textdaggerdbl}  & 1.00   & 0.81     & 0.09       & Glitch     & False                     \\
S191117j\textsuperscript{\textdagger}  & 1.00   & 0.00     & 0.02       & Glitch     & False                     \\
S191120at\textsuperscript{\textdagger} & 1.00   & 0.16     & 0.08       & Glitch     & True                      \\
S191212q\textsuperscript{\textdagger \textdaggerdbl}  & 1.00   & 0.90     & 0.69       & Glitch     & True                      \\
S191213g\textsuperscript{\textdagger \textdaggerdbl}  & 1.00   & 0.79     & 0.16       & Glitch     & False                     \\
S191220af\textsuperscript{\textdagger \textdaggerdbl \textasteriskcentered} & 1.00   & 0.84     & 0.77       & Glitch     & False                     \\
S191225aq\textsuperscript{\textdagger \textdaggerdbl \textasteriskcentered} & 1.00   & 1.00     & 0.96       & Glitch     & True                      \\
S200105ae\textsuperscript{\textdagger} & 1.00   & 0.35     & 0.42       & Glitch     & False                     \\
S200106au\textsuperscript{\textdagger \textdaggerdbl \textasteriskcentered} & 1.00   & 1.00      & 0.99       & Glitch     & True                      \\
S200106av\textsuperscript{\textdagger \textdaggerdbl \textasteriskcentered} & 1.00   & 0.90     & 0.89       & Glitch     & True                      \\
S200112r\textsuperscript{\textdaggerdbl \textasteriskcentered}  & 0.92   & 0.53     & 0.52       & Real       & False                     \\
S200116ah\textsuperscript{\textdagger} & 1.00   & 0.00     & 0.02       & Glitch     & False                     \\
S200302c\textsuperscript{\textdagger \textdaggerdbl \textasteriskcentered}  & 0.00   & 0.30     & 0.26       & Real       & False                     \\
S200308e\textsuperscript{\textdagger}  & 1.00   & 0.18     & 0.07       & Glitch     & True   \\                  
\hline 
\end{tabularx}
\caption{Misidentifications of O3 public alerts by the three \texttt{GWSkyNet} models. The first column displays the IDs of the alerts. The superscripts \textdagger, \textdaggerdbl ~and \textasteriskcentered indicate misidentifications by \cite{cabero2020gwskynet}, the Baseline model and the Fine-tuned model respectively. Columns 2-4 show the \texttt{GWSkyNet} scores for the alerts by the specific models. The true labels of the alerts are based on \cite{2021arXiv211103606T,LIGOScientific:2021usb}. The last column indicates whether the \ac{SNR} of the alerts would have met the \ac{SNR} requirement defined in this paper and thus whether would have been analyzed by \texttt{GWSkyNet}.\label{table:o3table}}
\end{table*}

\end{document}